\documentclass[lettersize,journal]{IEEEtran}
\usepackage{amsmath,amsfonts}
\usepackage{algorithmic}
\usepackage{array}
\usepackage[caption=false,font=normalsize,labelfont=sf,textfont=sf]{subfig}
\usepackage{textcomp}
\usepackage{stfloats}
\usepackage{url}
\usepackage{verbatim}
\usepackage{graphicx}
\usepackage{cite}
\usepackage[colorlinks=true,linkcolor=blue,citecolor=blue,urlcolor=blue]{hyperref}
\usepackage{booktabs}
\usepackage{tabularx}
\usepackage{multirow}
\usepackage{amssymb}
\def\BibTeX{{\rm B\kern-.05em{\sc i\kern-.025em b}\kern-.08em
    T\kern-.1667em\lower.7ex\hbox{E}\kern-.125emX}}
\usepackage{balance}
\begin{document}
\title{Frequency Response of Windowed DFT Phasor Estimation: Impact on Oscillation Observability}
\author{Jiahui Yang, \textit{Student Member}, \textit{IEEE}, Yuru Wu, \textit{Member}, \textit{IEEE}, Haozong Wang,

Yu Liu, Biao Sun, Yilu Liu, \textit{Fellow}, \textit{IEEE}, Clifton Black, \textit{Member}, \textit{IEEE}
\thanks{This work is partially supported by the CURENT Industry Partnership Program. (Corresponding Author: Yuru Wu(ywu70@vols.utk.edu))

}}


\maketitle

\begin{abstract}
Phasor measurement units (PMUs) are widely used for sub-synchronous oscillation monitoring, yet the effect of windowed discrete Fourier transform (DFT)-based phasor estimation on oscillation observability is not fully characterized. This letter derives the complete complex-valued frequency response of the windowed DFT phasor estimator under both magnitude and phase modulation. The analysis shows that the estimation window introduces both frequency-dependent magnitude attenuation and phase shift to oscillation components, governed by the complex gain. A simple recovery method is also proposed to restore the true oscillation amplitude and phase from PMU data using the analytically known complex gain. The results are validated through time-domain simulations and provide guidance for industry practitioners on interpreting PMU-based oscillation measurements and selecting appropriate window lengths.

\end{abstract}


\vspace{-1 em}
\section{Introduction}
\IEEEPARstart{T}{he} increasing penetration of inverter-based resources (IBRs) in power systems has led to sub-synchronous oscillations spanning a broad frequency range (e.g., 0.1--30\,Hz) \cite{cheng2022real}. Such oscillations can degrade equipment life and may even precipitate blackouts if sustained. This backdrop makes oscillation monitoring and detection critical for maintaining reliability in high-renewable power grids. 

Phasor measurement units (PMUs) \cite{yang2023real} are widely used for oscillation monitoring due to the high reporting rate, synchronized measurements, and wide geographic coverage \cite{ma2021application}, but most detection methods implicitly assume that PMU measurements faithfully preserve oscillatory components. In practice, most PMU outputs are processed through windowed discrete Fourier transform (DFT)-based phasor estimation, and the choice of window length---short for P-class and long for M-class devices per IEEE Standard C37.118.1 \cite{C37.118}---leads to different frequency responses that can distort oscillations.

Recent work has begun to address this issue. Follum \textit{et al.} \cite{follum2021phasors} highlight the impact of PMU preprocessing and anti-aliasing filters on high-frequency dynamics. Ou \textit{et al.} \cite{ou2026applicability} analyze DFT-induced amplitude attenuation and identify a ``phase flip'' when the measurement gain changes sign, though only magnitude modulation is considered in their signal model.

This letter addresses this gap by deriving the complete complex-valued frequency response of the windowed DFT phasor estimator under both magnitude and phase modulation, quantifying the frequency-dependent magnitude attenuation and phase shift imposed on oscillation components. Understanding these effects is critical for grid operators, since ignoring the frequency response may cause oscillations to be missed entirely or their severity to be underestimated. Based on the derived response, a recovery method is proposed to restore the true oscillation parameters from PMU data. The results provide practical guidance on selecting PMU window lengths for sub-synchronous oscillation detection.
\vspace{-0.5 em}
\section{Windowed DFT Phasor Estimation Model}
\label{sec:method}

\subsection{Signal Models for Modulation}
\label{subsec:signal_model}
A single-phase sinusoidal waveform subject to either magnitude or phase modulation is considered, respectively \cite{C37.118}, which are sufficient to capture the two most common oscillations. 

a) Magnitude modulation is modeled as in Eq. \eqref{eq:mm_wave}:
\begin{gather}
x[p] = \sqrt{2}V_{\mathrm{rms}}\bigl(1+\alpha \sin(\Omega_m p+\phi_m)\bigr)\cos(\omega_0 p+\phi_0)
\label{eq:mm_wave}
\end{gather}
where $V_{\mathrm{rms}}$ is the RMS magnitude, and $\alpha$ is the magnitude-modulation index. $p$ is the discrete-time sample index. $\phi_m$ and $\phi_0$ denote the initial carrier phase and oscillation phase. $\Omega_m = 2\pi f_m/f_s$. $\omega_0 = 2\pi f_0/f_s$. $f_0$ and $f_m$ are the carrier frequency and oscillation frequency, $f_s$ is the sampling rate.

b) Phase modulation is modeled as in Eq. \eqref{eq:pm_wave}.
\begin{equation}
x[p] = \sqrt{2}V_{\mathrm{rms}}\cos\!\bigl(\omega_0 p+\phi_0 + \beta \sin(\Omega_m p+\phi_m)\bigr)
\label{eq:pm_wave}
\end{equation}
where $\beta$ is the phase-modulation index.
\vspace*{-1 em}
\subsection{Windowed DFT Phasor Estimation}
\label{subsec:wdft}
Phasor estimation is commonly based on multi-cycle DFT:
\begin{gather}
X_m(k)=\frac{\sqrt{2}}{L}\sum_{n=0}^{L-1} x[m+n]e^{-j\omega_k (m+n)} \label{eq:wdft_def}
\end{gather}
where $L=hN$ is the window length in samples ($h$ cycles, $N$ samples/cycle). $m$ is the frame index. $k$ is the DFT bin index.  $\omega_k=2\pi k/L$. The corresponding bin frequency $f_k = k f_s/L$. For synchrophasor estimation, $f_k=f_0$. Hence, $h=k$ and $\omega_k=\omega_0$. Eq. \eqref{eq:wdft_def} can be written in equivalent form:
\begin{gather}
y[p] = x[p]e^{-j\omega_0 p}
\label{eq:demod_def} \\
\hat{X}[m] = \frac{\sqrt{2}}{L}\sum_{n=0}^{L-1}y[m+n]
\label{eq:phasor_est}
\end{gather}
where $y[p]$ denotes the demodulated signal obtained by shifting $x[p]$ to the synchronous rotating reference frame at the nominal angular frequency.
\vspace{-1 em}
\subsection{Decomposition of Demodulated Signal \texorpdfstring{$y[p]$}{y[p]}}
\label{subsec:baseband_detail}
This subsection analyzes the demodulated signal $y[p]$ under both magnitude and phase modulation, respectively. 

a) Magnitude modulation: Using Euler's formulas:
\begin{gather}
\cos(\omega_0 p + \phi_0) = \frac{1}{2}\left(e^{j(\omega_0 p + \phi_0)} + e^{-j(\omega_0 p + \phi_0)}\right) 
\label{eq:Euler_cos}
\end{gather}
\begin{gather}
\sin(\Omega_m p + \phi_m) = \frac{1}{2j}\left(e^{j(\Omega_m p + \phi_m)} - e^{-j(\Omega_m p + \phi_m)}\right) \label{eq:Euler_sin}
\end{gather}
Substituting Eq. \eqref{eq:mm_wave}, \eqref{eq:Euler_cos} and \eqref{eq:Euler_sin} into Eq. \eqref{eq:demod_def}:
\begin{gather}
y[p] = \frac{\sqrt{2}V_{\mathrm{rms}}}{2}\left(e^{j\phi_0} + e^{-j(2\omega_0 p + \phi_0)}\right) +\frac{\sqrt{2}V_{\mathrm{rms}}\alpha}{4j} \nonumber\\
\left(e^{j(\Omega_m p + \phi_0 + \phi_m)} - e^{-j(\Omega_m p - \phi_0 + \phi_m)}\right) +\frac{\sqrt{2}V_{\mathrm{rms}}\alpha}{4j} \nonumber \\
\left(-e^{-j((2\omega_0 + \Omega_m) p + \phi_0 + \phi_m)} + e^{-j((2\omega_0 - \Omega_m) p + \phi_0 - \phi_m)}\right)
\label{eq:yp_mag}
\end{gather}

b) Phase modulation: Similarly, using Euler's formula:
\begin{gather}
y[p] = \frac{\sqrt{2}V_{\mathrm{rms}}}{2}(e^{j(\phi_0 + \beta\sin(\Omega_m p + \phi_m))} \nonumber \\ 
+ e^{-j(2\omega_0 p + \phi_0 + \beta\sin(\Omega_m p + \phi_m))} )
\end{gather}
Compared with the base signal, the oscillation index $\beta$ is small. Then, using $e^{j\beta\sin\theta} \approx 1 + j\beta\sin\theta$ (the error is below 0.5\% if $\beta<0.1$ rad\cite{C37.118}) and applying Euler's formula:
\begin{gather}
y[p] = \frac{\sqrt{2}V_{\mathrm{rms}}}{2}\left(e^{j\phi_0} + e^{-j(2\omega_0 p + \phi_0)}\right) + \frac{\sqrt{2}V_{\mathrm{rms}}\beta}{4} \nonumber \\
 \left(e^{j(\Omega_m p + \phi_0 + \phi_m)} - e^{-j(\Omega_m p - \phi_0 + \phi_m)}\right) + \frac{\sqrt{2}V_{\mathrm{rms}}\beta}{4} \nonumber \\
 \left(-e^{-j((2\omega_0 + \Omega_m)p + \phi_0 + \phi_m)} + e^{-j((2\omega_0 - \Omega_m)p + \phi_0 - \phi_m)}\right)
\label{eq:phase_mod_grouped}
\end{gather}

So, for both magnitude and phase modulation, the demodulated signal $y[p]$ can be decomposed into a DC component and five exponential terms rotating at angular frequencies: $-2\omega_0, \pm \Omega_m$, and $-2\omega_0 \pm \Omega_m$.

\vspace{-1 em}
\subsection{Frequency Response of Windowed DFT}
Since $y[p]$ is a sum of complex exponentials at different frequencies, the estimator output is the sum of individual responses. The response to each term can be analyzed by substituting a general test signal $y[m] = e^{j\lambda m}$ into Eq.~\eqref{eq:phasor_est} and evaluating the resulting gain at each $\lambda$.
\begin{gather}
\hat{X}[m] = \frac{\sqrt{2}}{L}\sum_{n=0}^{L-1} e^{j\lambda(m+n)} = e^{j\lambda m} H_1(e^{j\lambda})
\label{eq:gain_lambda}\\
H_1(e^{j\lambda}) = \frac{\sqrt{2}}{L}\sum_{n=0}^{L-1} e^{j\lambda n} = \frac{\sqrt{2}}{L}e^{j\lambda(L-1)/2} \frac{\sin(L\lambda/2)}{\sin(\lambda/2)}
\label{eq:freq_response}
\end{gather}
\begin{figure}[!b]
  \centering
  \includegraphics[width=\linewidth]{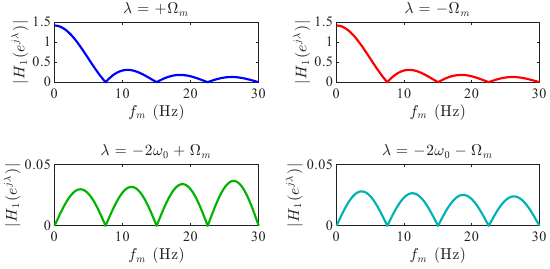}
  \caption{Magnitude of the complex gain $H_1$ versus oscillation frequency $f_m$.}
  \label{fig:H_Mag}
\end{figure}

From Eq.~\eqref{eq:gain_lambda}, the output for each exponential term is the input scaled by $H_1$, which therefore serves as the complex gain of the windowed DFT. Evaluating Eq.~\eqref{eq:freq_response} at each frequency: for the DC term, $\lambda_1 = 0$ and $|H_1| \triangleq \sqrt{2}$; for $\lambda_2 = -2\omega_0$, $|H_1| \triangleq 0$. For the remaining terms $\lambda_{3,4} = \pm \Omega_m$ and $\lambda_{5,6}=-2\omega_0 \pm \Omega_m$, Fig.~\ref{fig:H_Mag} shows their magnitude under $h = 8$, where $\lambda_{5,6}$ are negligibly small. $\lambda_{5,6}$ arise from the demodulation, not physical oscillations; for smaller $h$, they grow but can be suppressed by a low-pass filter prior to downsampling, ensuring accurate oscillation estimation without aliasing.

By neglecting the $\lambda_{5,6}$ terms, $y[p]$ reduces to DC components and exponential terms at $\lambda_{3,4}$. For magnitude modulation, substituting the remaining elements of $y[p]$ from Eq. \eqref{eq:yp_mag} into Eq.~\eqref{eq:gain_lambda} separately and summing their responses:
\begin{gather}
\hat{X}[m] = \frac{\sqrt{2}V_{\mathrm{rms}}\alpha}{4j}e^{j(\phi_0 + \phi_m)} H_1(e^{j\Omega_m}) e^{j\Omega_m m} - \frac{\sqrt{2}V_{\mathrm{rms}}\alpha}{4j} \cdot \nonumber\\  e^{j(\phi_0 - \phi_m)} H_1(e^{-j\Omega_m}) e^{-j\Omega_m m} + C_0
\label{eq:x_mod_mag}
\end{gather}
where $C_0 = V_{\mathrm{rms}}e^{j\phi_0}$. Let $H_1(e^{j\Omega_m}) = G \cdot e^{j\theta}$, $G = |H_1(e^{j\Omega_m})|$ and $\theta = \angle H_1(e^{j\Omega_m})$. Due to $H_1(e^{-j\lambda}) = H_1^*(e^{j\lambda})$, $H_1(e^{-j\Omega_m}) = G \cdot e^{-j\theta}$. Eq. \eqref{eq:x_mod_mag} can be written:  
\begin{gather}
\hat{X}[m] = G \frac{\sqrt{2}V_{\mathrm{rms}}\alpha}{2}  e^{j\phi_0} \sin(\Omega_m m + \phi_m + \theta) + C_0
\label{eq:mag_xm}
\end{gather}

Similarly, for phase modulation, 
\begin{gather}
\hat{X}[m] = G \frac{\sqrt{2}V_{\mathrm{rms}}\beta}{2} j\sin(\Omega_m m + \phi_m + \theta) e^{j\phi_0}+C_0
\end{gather}
Applying $e^{j\beta\sin\theta} \approx 1 + j\beta\sin\theta$ again:
\begin{equation}
\hat{X}[m] = V_{\mathrm{rms}}e^{j\left(\phi_0 + G  \frac{\sqrt{2}\beta}{2} \sin(\Omega_m m + \phi_m + \theta)\right)}
\label{eq:pha_xm}
\end{equation}

Eq. \eqref{eq:mag_xm} and \eqref{eq:pha_xm} show that for both modulation types the windowed DFT affects the modulation in two ways: (i) magnitude scaled by $G=\left|H_1\!\left(e^{j\Omega_m}\right)\right|$; and (ii) phase shifted by $\theta=\angle H_1\!\left(e^{j\Omega_m}\right)=\Omega_m(L-1)/2$ (with an additional $\pi$ offset when the real gain changes sign). 

\begin{figure}[!t]
  \centering
  \includegraphics[width=\linewidth]{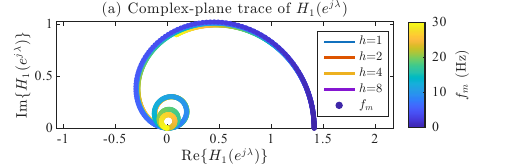}
  \vspace{-1mm} 
  \includegraphics[width=\linewidth]{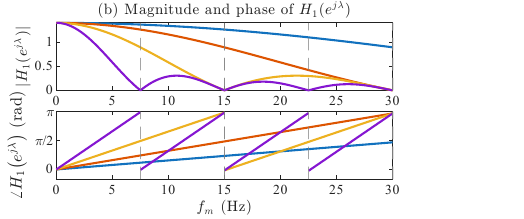}
  \captionsetup{skip=-2pt} 
  \caption{Complex-plane trajectory, magnitude, and phase of $H_1(e^{j\lambda})$.}
  \label{fig:H_compex_m_p}
\end{figure}

Fig.~\ref{fig:H_compex_m_p} shows the complex-plane trace, magnitude, and phase of $H_1$ for different $h$ and $f_m$. For small $f_m$, windowing has little effect ($G\approx \sqrt{2}$, $\theta$ is small). As $f_m$ increases, attenuation grows with $h$, and periodic nulls completely cancel oscillatory components. The phase varies continuously with frequency. 

Note that the windowed DFT output is naturally referenced to the left edge $m$, used in Eq. \eqref{eq:wdft_def}. For a pure sinusoid, the phasor is independent of timestamp convention, but for oscillation components, adopting a center timestamp absorbs the phase shift $\theta$ into the time coordinate, reducing the gain to a real value and masking the frequency-dependent phase distortion \cite{ou2026applicability}. Nevertheless, center-of-window timestamp also preserves relative oscillation phase differences across PMUs.

\vspace{-0.5 em}
\subsection{Oscillation Recovery}
Once an oscillation is detected in PMU data, its frequency $f_m$ can be estimated via spectral analysis. With the known window length $h$, the complex gain $H_1(e^{j\lambda})$ is fully determined. Provided $G \neq 0$ (i.e., $f_m$ does not fall on a comb-null), the true oscillation amplitude and phase can be recovered as $A_{\mathrm{rec}} = \sqrt{2}A_{\mathrm{meas}}/G$ and $\phi_{\mathrm{rec}} = \phi_{\mathrm{meas}} - \theta$, enabling accurate assessment of oscillation severity.

\section{Case Studies}
To validate the proposed model, time-domain simulations were conducted with a sampling rate of 960~Hz, a reporting rate of 60~fps, $V_{\mathrm{rms}}=1.0$, and window lengths $h=1,2,4,8$. The windowed DFT is applied to estimate phasor, and the oscillation is extracted via least-squares sinusoidal fitting and compared with the theoretical predictions.

\begin{figure}[!t]
  \centering
  \includegraphics[width=\linewidth]{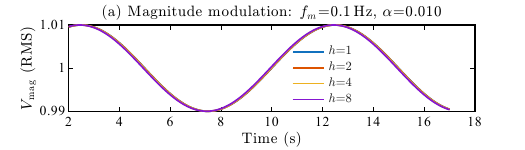}
  \vspace{-1mm} 
  \includegraphics[width=\linewidth]{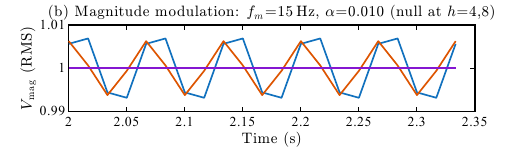}
  \vspace{-1mm} 
  \includegraphics[width=\linewidth]{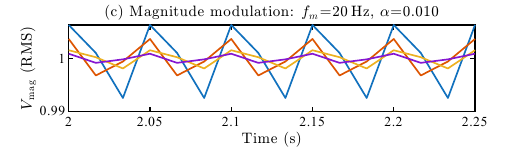}
  \vspace{-1mm} 
  \includegraphics[width=\linewidth]{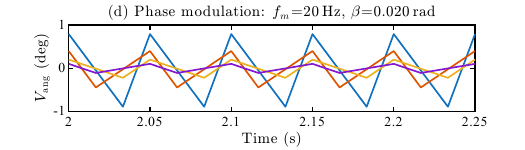}
  \captionsetup{skip=-4pt} 
  \caption{Phasor data using windowed DFT at 60 fps.}
  \label{fig:case study}
\end{figure}

Fig.~\ref{fig:case study} shows the phasor outputs under four scenarios across magnitude and phase modulation. At $f_m=0.1$\,Hz, all window lengths produce nearly identical outputs with negligible attenuation, consistent with Fig.~\ref{fig:H_compex_m_p} for small $f_m$. At $f_m=15$\,Hz, the oscillation is completely suppressed for $h=4$ and $h=8$ due to the comb-null condition in Fig.~\ref{fig:H_compex_m_p}. At $f_m=20$\,Hz, progressive attenuation with increasing $h$ is clearly visible in both the magnitude and angle modulation, and phase shifts between different $h$ values are evident. 

Table~\ref{tab:validation} compares the measured oscillation magnitude and phase with the theoretical values from Eq.~\eqref{eq:mag_xm} and \eqref{eq:pha_xm}. The magnitude measurements confirm the predicted gain for each $h$, and the phase measurements are consistent with $\theta$ up to a small offset caused by aliasing of $-2\omega_0 \pm \Omega_m$ image terms onto $f_m$ at 60 fps. At 240 fps, these terms no longer alias onto $f_m$, and the $h=8^*$ column shows near-perfect agreement with theory; alternatively, a low-pass filter prior to downsampling to 60 fps can suppress them. These results validate that the windowed DFT introduces both frequency-dependent magnitude attenuation and phase shift to oscillation components, and the close agreement between theory and measurement confirms that $H_1$ can reliably recover the true oscillation from PMU data if it does not fall on a comb-null.
\vspace{-0.8 em}
\section{Conclusion}
\vspace{-0.2 em}
This letter derives the complete complex-valued frequency response of the windowed DFT phasor estimator and shows that it introduces both frequency-dependent magnitude attenuation and phase shift to oscillation components. Since $H_1$ is analytically known, the true oscillation amplitude and phase can be recovered from PMU measurements. For practical applications: the absence of oscillations in phasor data does not guarantee oscillation-free operation, as it may be canceled at comb-null frequencies or heavily attenuated. Furthermore, detected oscillation magnitudes may underestimate the true severity. Therefore, shorter DFT windows are preferable for sub-synchronous oscillation monitoring, while an anti-aliasing filter is recommended to suppress image components.
\begin{table}[!t]
\begin{center}
\caption{Comparison of Oscillation Amplitude and Phase at $f_m=20$\,\textnormal{Hz}. \textnormal{(1) Magnitude modulation (amplitude in RMS, phase in degrees). (2) Phase modulation (amplitude and phase in degrees).}}
\label{tab:validation}
\begin{tabularx}{\columnwidth}{c c >{\centering\arraybackslash}X >{\centering\arraybackslash}X >{\centering\arraybackslash}X >{\centering\arraybackslash}X >{\centering\arraybackslash}X}
\toprule
 & & $h=1$ & $h=2$ & $h=4$ & $h=8$ & $h=8^*$ \\
\midrule
\multirow{4}{*}{(1)} & $A_{\mathrm{theory}}$  & 0.008276 & 0.004138 & 0.002069 & 0.001034 & 0.001034 \\
 & $A_{\mathrm{meas}}$    & 0.008045 & 0.004016 & 0.002011 & 0.001004 & 0.001034 \\
 & $\theta_{\mathrm{theory}}$ & 56.25 & 116.25 & 56.25 & 116.25 & 116.25 \\
 & $\theta_{\mathrm{meas}}$   & 52.09  & 112.45  & 52.10  & 112.45 & 116.26 \\
\midrule
\multirow{4}{*}{(2)} & $A_{\mathrm{theory}}$  & 0.9483 & 0.4742 & 0.2371 & 0.1185 & 0.1185 \\
 & $A_{\mathrm{meas}}$    & 0.9673 & 0.4846 & 0.2421 & 0.1211 & 0.1185 \\
 & $\theta_{\mathrm{theory}}$ & 56.25 & 116.25 & 56.25 & 116.25 & 116.25 \\
 & $\theta_{\mathrm{meas}}$   & 59.82  & 120.18  & 59.82  & 120.18 & 116.27 \\
\bottomrule
\multicolumn{7}{l}{\footnotesize $^*$240\,fps reporting rate; all other columns use 60\,fps.}
\end{tabularx}
\end{center}
\end{table}
\vspace{-1em}
\bibliographystyle{IEEEtran}
\bibliography{ref}

\end{document}